# Effective dislocation lines in continuously dislocated crystals
## III. Kinematics

ANDRZEJ TRZĘSOWSKI

**Institute of Fundamental Technological Research**
**Polish Academy of Sciences**
Świętokrzyska 21, 00-049 Warsaw, Poland
e-mail: atrzes@ippt.gov.pl or atrzes@wp.pl

A class of congrunces of principal Volterra-type effective dislocation lines associated with a dislocation density tensor is distinguished in order to investigate the kinematics of continuized defective crystals in terms of their dislocation densities (tensorial as well as scalar). Moreover, it is shown , basing oneself on a formula defining the mean curvature of glide surfaces for principal edge effective dislocation lines, that the considered kinematics of continuized defective crystals is consistent with some relations appearing in the physical theory of plasticity (e.g. with the Orowan-type kinematic relations and with treatment of the shear stresses as driving stresses of moving dislocations).

## 1. Introduction

If the macroscopic properties of a crystalline solid with many dislocations are considered, a *continuous limit approximation* can be defined by means of the condition that, at each point of the body, a *characteristic mesoscopic length*, say of the order of 10-100 nm, can be approximately replaced with the infinitesimal length [1]. Although, in this continuous limit, the global long-range order of crystals is lost in the presence of dislocations, nevertheless their local long-range order still exists [2]. We restrict our investigation to the *Bravais crystal*, because this crystal has the smallest amount of different defect types, but enough to study the general principles.

The mesoscopic observation level scale enables to consider the so-called *mesoplasticity* approach to the description of plastic deformations [3]. In the mesoplasticity approach, plastic deformation can be, at least in principle, predicted by an Orowan-type theoretical model, that is, by generalization of the so-called *Orowan kinematic relation* [3, 4]:

(1.1) $$\dot{\gamma} = \rho b v, \qquad [\dot{\gamma}] = \text{s}^{-1}, \qquad [v] = \text{cms}^{-1},$$

where $\dot{\gamma}$ denotes the macroscopic strain rate, $\rho$ is the mean density of mobile dislocations defined as the length of all moving dislocation lines included in the volume unit, $b$ is the mean strength of these dislocations (i.e. the mean modulus of their Burgers vectors), and $v$ is the mean dislocation speed. There are two basic types of dislocation movement, *glide* in which the dislocation moves in a surface ,called the *glide surface*, which contains its line and Burgers vector, and *climb* in which the dislocation moves out of the glide surface normal to the Burgers vector [4]. Glide of many dislocations results in *slip*, which is the most common manifestation of plastic deformation in crystalline solids. The glide motion of an *effective dislocation line* ([6] and Section 3) can be considered as an elementary act of the mesoscale-type continuous limit description of plastic deformations [5, 6]. It is shown that the geometric theory of continuously dislocated crystals presented in [1] and [6] admits a continuous counterpart of the Orowan kinematic relation (1.1) (Sections 4-6 and [7]).

The appearance of dislocations generates a bend of originally straight lattice lines. For example, the *lattice lines* in a continuized dislocated Bravais crystal form a system of three independent congruences of curves and tangents to these curves define *local crystallographic directions* of this crystal. Planes spanned by two local crystallographic directions are *local crystal planes*. If a distribution of these planes is integrable [1], then its integral manifolds constitute a family of *crystal surfaces*. Note that in the case of a crystalline body with many dislocations, the mean value $\bar{\kappa}_n$ of normal curvatures of its crystal surfaces in their local crystallographic directions (see e.g. [1, 8]) can be approximated (in a continuous limit) by

(1.2) $\qquad \bar{\kappa}_n = \rho b, \qquad [\bar{\kappa}_n] = \text{cm}^{-1}, \qquad [\rho] = \text{cm}^{-2}, \qquad [b] = \text{cm},$

where $\rho$ denotes the mean density of dislocations defined here as the length of all dislocation lines included in the volume unit, and $b$ is the mean strength of the considered dislocations [6, 9]. It appears that the proposed theory of continuously dislocated crystals is consistent with this relation (Section 6; see also [6])

It is known that the occurrence of many dislocations in a crystalline solid is accompanied with the appearance of point defects created by the distribution of dislocations [10]. The influence of these *secondary point defects* on metric properties of a continuously dislocated Bravais crystal can be represented by a *Riemannian material space*, defined by the assumption that the body under consideration is additionally endowed with the such Riemannian *internal length measurement* that reduces to the Euclidean length measurement if the dislocations are absent (Section 2) [1]. The influence of secondary point defects on the slip phenomenon can be then taken into account by means of treatment of the congruences of effective dislocation lines, crystal surfaces and the virtual slip surfaces as those located in this material space (Sections 3 and 5). Note that such surfaces can be, at least locally, isometrically embedded in the Euclidean ambient space of the dislocated crystalline body [1]. Moreover, if a counterpart of Eq. (1.2) holds in the Riemannian material space, then the *mean curvature* of crystal surfaces, considered as submanifolds of this space, takes the physical meaning of a material parameter that characterizes the influence of secondary point defects on the slip phenomenon (Sections 6 and 7).

## 2. Dislocation densities

Let $B \subset E^3$ denote a *body* identified with its distinguished spatial configuration being an open and contractible to a point subset of the three-dimensional Euclidean point space $E^3$ (e.g. [2]). We will consider the curvilinear coordinate systems $X = (X^A; A = 1,2,3)$ defined on an open subset $U \subset B$ and such that $[X^A] = \text{cm}$ and we will denote $X = X(p) \in \mathbb{R}^3$ for $p \in U$. The body under consideration is a continuous *solid body* with its material structure defined as a continuous limit approximation of a Bravais crystal with many dislocations (see Section 1). A distinguished vector base $\Phi = (\mathbf{E}_a; a = 1,2,3)$ of the linear module $W(B)$ of vector fields on $B$ tangent to $B$ (see [1], Appendix), called further on the *Bravais moving frame*, is considered as the one defining a system of three independent congruences of *lattice lines* of the continuized crystal as well as scales of an *internal length measurement* along these lines [1]. The condition that the bend of lattice lines due to dislocations (see Section 1) is not generated by a global deformation of the body means that the so-called *object of anholonomity* $C_{ab}^c \in C^\infty(B)$ defined by (see [1], Appendix)

$$(2.1) \qquad [\mathbf{E}_a, \mathbf{E}_b] = C_{ab}^c \mathbf{E}_c,$$

does not vanish. This object of *material anholonomity* describes the *long-range distortion* of a Bravais crystal due to dislocations [1].

Let us denote by $W^*(B)$ the linear module of covector fields on $B$ dual to $W(B)$ [1]. The vectorial base $\Phi^* = (E^a)$ of $W^*(B)$ dual to the moving frame $\Phi = (\mathbf{E}_a)$ and called further on a *Bravais moving coframe*, is uniquely defined by the condition:

$$(2.2) \qquad \mathbf{E}_a = e_a^A \partial_A, \quad E^a = e^a_A dX^A \quad \Rightarrow \quad \langle E^a, \mathbf{E}_b \rangle = e^a_A e_b^A = \delta_b^a,$$

where, according to the assumed dimensional convention, we have:

(2.3) $\qquad [\mathbf{E}_a] = [\partial_A] = \text{cm}^{-1}, \qquad [\mathrm{E}^a] = [d\mathrm{X}^A] = \text{cm}.$

We can define the tensorial representation $\mathbf{S}[\Phi]$ of the object of anholonomity. Namely, if $\Phi = (\mathbf{E}_a)$ is a Bravais moving frame and $\Phi^* = (\mathrm{E}^a)$ is the *Bravais moving coframe* dual to $\Phi$, then we define [1, 11]:

(2.4) $\qquad \begin{aligned} \mathbf{S}[\Phi] &= d\mathrm{E}^a \otimes \mathbf{E}_a = S_{ab}{}^c \mathrm{E}^a \otimes \mathrm{E}^b \otimes \mathbf{E}_c, \\ S_{ab}{}^c &= -\frac{1}{2} C_{ab}^c, \qquad [S_{ab}{}^c] = \text{cm}^{-1}. \end{aligned}$

$\mathbf{S}[\Phi]$ characterizes the existence of many dislocations in this sense that

(2.5) $\qquad \mathbf{S}[\Phi] = \mathbf{0} \quad \text{iff} \quad \mathbf{E}_a \doteq \partial/\partial \xi^a, \quad a = 1,2,3,$

where $\xi = (\xi^a), [\xi^a] = \text{cm}$, is a coordinate system on $B$ and $\doteq$ means that a relation is valid in a distinguished coordinate system. Thus, the tensor field $\mathbf{S}[\Phi]$ can be interpreted as a *nondimensional measure* of the long-range distortion of the dislocated Bravais crystal due to bending of originally straight lattice lines [1]. This long-range distortion of the dislocated Bravais crystal can be quantitatively measured by the so-called *Burgers vector* $\mathbf{b}[\gamma]$, corresponding to a closed smooth contour $\gamma$ (called a *Burgers circuit*) in the considered defective crystalline solid body $B$ [1]:

(2.6) $\qquad \begin{aligned} \mathbf{b}[\gamma] &= b^a[\gamma]\mathbf{C}_a, \qquad b^a[\gamma] = \varepsilon \oint_\gamma \mathrm{E}^a, \\ [\mathbf{b}[\gamma]] &= 1, \qquad [b^a[\gamma]] = \text{cm}, \qquad \varepsilon = \pm 1, \end{aligned}$

where $\varepsilon$ defines the Burgers vector orientation and $C = (\mathbf{C}_a; a = 1,2,3)$ is an orthonormal Cartesian base of the Euclidean vector space $\mathbf{E}^3$ of translations in $E^3$.

It seems to be physically reasonable to take into account the influence of secondary point defects on the Burgers vector. It can be done e.g. in the following way.

Firstly, let us note that although translational symmetries of the crystal are lost in the above mentioned continuous limit approximation, nevertheless the base vector fields of a Bravais moving frame can be considered as those that define the scales of an internal length measurement along local crystallographic directions of the dislocated Bravais crystal (Section 1). Namely, we can define the following *intrinsic material metric tensor* **g** of an *internal length measurement* within the dislocated Bravais crystal [1]:

$$\mathbf{g} = \mathbf{g}[\Phi] = \delta_{ab} \mathrm{E}^a \otimes \mathrm{E}^b = g_{AB} dX^A \otimes dX^B,$$

(2.7)

$$g_{AB} = \delta_{ab} e^a_A e^b_B, \qquad [\mathbf{g}] = \mathrm{cm}^2,$$

describing a distortion of the globally Euclidean length measurement within a crystalline body $B$ (embedded in its configurational Euclidean point space $E^3$) due to many dislocations. Since the Riemannian metric is locally Euclidean, therefore it is an internal length measurement consistent with the observed phenomenon that dislocations have no influence on the local metric properties of the crystalline body. Next, let us consider a Burgers circuit $\gamma \subset B$ as the one located in the *Riemannian material space* $B_g = (B, \mathbf{g})$, where **g** is the intrinsic metric tensor defined by Eq. (2.7). Then, the integrals of Eq. (2.6) that define components $\mathrm{b}^a[\gamma]$, $a = 1, 2, 3$, of the Burgers vector $\mathbf{b}[\gamma]$, can be treated as functionals in the Riemannian space $B_g$, defining a mapping $\gamma \subset B_g \to \mathbf{b}[\gamma] \in E^3$. Let $\Sigma \subset B$ be a surface possessing the closed contour $\gamma$ as its boundary and treated as a two-dimensional compact, connected and oriented Riemannian submanifold of $B_g$. Since

(2.8)
$$\mathrm{b}^a[\gamma] = \varepsilon \int_\Sigma dE^a,$$

it follows from the Stokes theorem in a Riemannian manifold [12] that the components $b^a[\gamma]$ of the Burgers vector $\mathbf{b}[\gamma]$ can be written in the following form [1]:

(2.9)
$$b^a[\gamma] = \int_\Sigma \alpha^{ba} d\Sigma_b, \qquad d\Sigma_b = d\Sigma l_b,$$
$$l_a = \delta_{ab} l^b, \qquad \|\mathbf{l}\|_g = l_a l^a = 1, \qquad \mathbf{l} = l^a \mathbf{E}_a,$$

where $\mathbf{l} \in W(B)$ is a unit vector field normal to the surface element $d\Sigma$ of $\Sigma$ and it was denoted

(2.10)
$$\alpha^{ba} = \varepsilon S_{cd}{}^a e^{cdb}, \qquad [\alpha^{ba}] = \text{cm}^{-1},$$

where and $e^{abc} \doteq \varepsilon^{abc}$ denotes the permutation symbol associated with the Bravais-moving frame $\Phi$ and considered as components in this base of a contravariant 3-vector density of weight +1 in $B_g$. The *dislocation density tensor* $\boldsymbol{\alpha}$ is defined as [1]:

(2.11)
$$\boldsymbol{\alpha} = \alpha^{ab} \mathbf{E}_a \otimes \mathbf{E}_b, \qquad [\boldsymbol{\alpha}] = \text{cm}^{-3}.$$

Likewise, the *scalar volume dislocation density* $\rho$ of a finite total length $L_d(B)$ of dislocation lines located in $B$ is defined by the following formula:

(2.12)
$$0 < L_d(B) = \int_B \rho \omega_g = \int_B \rho dV_g < \infty, \qquad [L_d(B)] = \text{cm}, \qquad [\rho] = \text{cm}^{-2},$$
$$\omega_g = \mathbf{E}^1 \wedge \mathbf{E}^2 \wedge \mathbf{E}^3 = e\, dX^1 \wedge dX^2 \wedge dE^3, \qquad dV_g = \sqrt{g}\, dX^1 dX^2 dX^3,$$
$$e = \det\left(e^a_A\right) = \sqrt{g}, \qquad g = \det(g_{AB}), \qquad [\omega_g] = \text{cm}^3,$$

where $\omega_g$ is the volume 3-form of $B_g = (B, \mathbf{g})$ defined by Eqs. (2.2) and (2.7) and $dV_g$ denotes the *material volume element.*

The components $\alpha^{ab}$ of the dislocation density tensor can be written in the following form [1]:

(2.20)
$$\alpha^{ab} = \gamma^{ab} + \sigma^{ab},$$
$$\gamma^{ab} = \alpha^{(ab)}, \qquad \sigma^{ab} = \alpha^{[ab]} = \frac{1}{2}t_c e^{cab},$$

where, according to Eqs. (2.4) and (2.10), we have:

(2.21)
$$t_a = e_{abc}\alpha^{bc} = \varepsilon C_{ab}^b, \qquad \varepsilon = \pm 1,$$

and $e_{abc} \doteq \varepsilon_{abc} (= \varepsilon^{abc})$ denote the permutation symbols associated with the Bravais moving coframe $\Phi^*$ and considered as components in this base of a covariant 3-vector density of weight $-1$ in $B_g$. It follows from Eqs. (2.4), (2.10), and (2.20) that

(2.22)
$$\varepsilon C_{ab}^c = t_{[a}\delta_{b]}^c - e_{abd}\gamma^{dc}.$$

Therefore, the long-range distortion of the continuously dislocated Bravais crystal with secondary point defects characterizes the following pair $(\gamma, \mathbf{t})$ defined in the Riemannian material space $B_g$ [1]:

(2.23)
$$\gamma = \gamma^{ab}\mathbf{E}_a \otimes \mathbf{E}_b, \qquad \gamma^{ab} = \gamma^{ba},$$
$$\mathbf{t} = t^a\mathbf{E}_a, \qquad t^a = \delta^{ab}t_b; \qquad \left[\gamma^{ab}\right] = \left[t^a\right] = \text{cm}^{-1}.$$

### 3. Effective dislocations

Let us rewrite Eq. (2.9) in the following form:

(3.1)
$$b^a[\gamma] = \int_\Sigma \rho b^a d\Sigma,$$
$$\rho b^a = l_b \alpha^{ba}, \qquad [b^a] = \text{cm},$$

where $\rho$ is the scalar density of dislocations defined by Eq. (2.12), and $C[\mathbf{l}]$ is a congruence in the material space $B_g$ defined by the unit vector field $\mathbf{l} = l^a\mathbf{E}_a$, $[\mathbf{l}] = \text{cm}^{-1}$, and by the following condition [6]:

(3.2) $$\rho\mathbf{b} = \mathbf{l}\alpha, \qquad \|\mathbf{l}\|_g = 1,$$

where

(3.3)
$$\mathbf{b} = b^a \mathbf{E}_a, \qquad [b^a] = \text{cm}, \qquad [\mathbf{b}] = 1;$$
$$b_g = \|\mathbf{b}\|_g = (b^a b_a)^{1/2} > 0, \qquad b_a = \delta_{ac} b^c, \qquad [b_g] = \text{cm}.$$

We will identify a geometric curve of this congruence with an *effective dislocation line* [6]. A line in $B_g$ with its unit tangent $\mathbf{l}$ and endowed with the nonvanishing local Burgers vector $\mathbf{b}$, can be interpreted as the *edge* (effective) dislocation line if [6]

(3.4) $$\mathbf{b} \cdot \mathbf{l} = b^a l_a = b_g m^a l_a = 0,$$

where it was denoted:

(3.5)
$$\mathbf{b} = b_g \mathbf{m}, \qquad b_g > 0,$$
$$\mathbf{m} = m^a \mathbf{E}_a, \qquad \|\mathbf{m}\|_g = (m^a m_a)^{1/2} = 1, \qquad [\mathbf{m}] = \text{cm}^{-1},$$

or - as the *screw* (effective) dislocation line if

(3.6) $$\mathbf{b} = \eta \mathbf{l}, \qquad \eta \neq 0, \qquad [\eta] = \text{cm}.$$

In other cases an effective dislocation line is interpreted as the *mixed* dislocation line.

Introducing the designations:

(3.7)
$$\cos\varphi_{l,t} = \frac{\mathbf{l} \cdot \mathbf{t}}{t_g}, \qquad 0 \leq \varphi_{l,t} \leq \pi,$$
$$t_g = \|\mathbf{t}\|_g = (t^a t_a)^{1/2}, \qquad [t_g] = \text{cm}^{-1},$$

and

(3.8)
$$\boldsymbol{\mu} = \mu^a \mathbf{E}_a = \mu \mathbf{m}, \qquad \|\mathbf{m}\|_g = 1,$$
$$\mu^a = \frac{1}{2} t_b l_c e^{bca}, \qquad \mu = \frac{1}{2} t_g \sin\varphi_{l,t} \geq 0,$$

we can write, according to Eqs. (2.11), (2.13), (3.2) and (3.8), the local Burgers vector $\mathbf{b}$ in the form [6]

$$\rho\mathbf{b} = \gamma\mathbf{l} + \mu\mathbf{m}, \qquad \mu \geq 0,$$
(3.9)
$$\|\mathbf{l}\|_g = \|\mathbf{m}\|_g = 1, \qquad \mathbf{l}\cdot\mathbf{m} = \mathbf{t}\cdot\mathbf{m} = 0,$$

where

(3.10) $$\rho\mathbf{b}\cdot\mathbf{l} = \mathbf{l}\gamma\mathbf{l} = \rho b_g \cos\varphi_{b,l}, \qquad \cos\varphi_{b,l} = \mathbf{b}\cdot\mathbf{l}/b_g.$$

For effective *screw* dislocation lines we have $\varphi_{b,l} = 0$ or $\pi$. For mixed or edge effective dislocation lines $\varphi_{b,l} \in (0, \pi)$. The family $\pi(\mathbf{l}, \mathbf{m})$ of planes spanned by the vector fields $\mathbf{l}$ and $\mathbf{m}$ constitute then an uniquely defined *two-dimensional distribution* on $B_g$ (see [1], Appendix) and the unit vector field $\mathbf{n}$ normal to these planes is uniquely defined up to its orientation. The planes of this distribution are *local slip planes* for the congruence $C[\mathbf{l}]$ if

(3.11) $$\mathbf{b}\cdot\mathbf{n} = 0,$$

or equivalently:

(3.12) $$\mathbf{n}\gamma\mathbf{l} = 0.$$

In this case $\mathbf{l}$ defines *locally Volterra-type* effective dislocation lines [6].

The ordered triple $\Upsilon = (\mathbf{l}, \mathbf{m}, \mathbf{n})$, defined by Eqs. (3.9)-(3.11) and the condition $\varphi_{b,l} \in (0, \pi)$, is called further on a *Volterra moving frame* [6]. It defines the two-dimensional *oriented distribution* $\pi_\mathbf{n}(\mathbf{l}, \mathbf{m})$ of local slip planes associated with the considered congruence of mixed (or edge) effective dislocation lines. If this oriented two-dimensional distribution is *integrable* ([12, 13]; see also [1], Appendix), then through each point of $B_g$ passes an oriented unique maximal integral manifold of the distribution. These integral manifolds are virtually *slip surfaces* (Section 1) for effective mixed dislocations of the considered congruence $C[\mathbf{l}]$ and the unit vector field $\mathbf{n}$ defines the congruence $C[\mathbf{n}]$ of curves normal to this family of (virtual) slip sur-faces. The Volterra moving frame $\Upsilon$ as well as the Bravais moving frame $\Phi$ span the linear module $W(B)$ of all smooth vector fields on $B$ tangent to $B$. Note that if

(3.13) $$\mathbf{n} = \mathbf{E}_3,$$

then the family $\pi_{\mathbf{n}}(\mathbf{l}, \mathbf{m})$ of local oriented slip planes defined by $\Upsilon$ is in coincidence with the distribution $\pi = \pi(\mathbf{E}_1, \mathbf{E}_2)$ of local crystal planes. In this case, the integral manifolds of the distribution $\pi$ define *crystal surfaces* being virtual *glide surfaces* for the considered congruence $C[\mathbf{l}]$ of Volterra-type effective dislocation lines [6].

Let us consider a *Frenet moving frame* $\Im = (\mathbf{e}_a; a=1,2,3)$ of vector fields on $B_g$ associated with the above-defined congruence $C[\mathbf{l}]$ of Volterra-type mixed effective dislocation lines, that is such that the *generalized formulae of Frenet* [6]

(3.14) $$\begin{aligned} \boldsymbol{\kappa} &= \nabla^g_{\mathbf{e}_1} \mathbf{e}_1 = \kappa \mathbf{e}_2, & \kappa &> 0, \\ \nabla^g_{\mathbf{e}_1} \mathbf{e}_2 &= -\kappa \mathbf{e}_1 + \tau \mathbf{e}_3, & \tau &\geq 0, \\ \nabla^g_{\mathbf{e}_1} \mathbf{e}_3 &= -\tau \mathbf{e}_2, & \kappa, \tau &\in C^\infty(B), \end{aligned}$$

where $\nabla^g$ denotes the Levi-Civita covariant derivative based on the Riemannian metric $\mathbf{g}$ (e.g. [14]), are valid. Moreover

(3.15) $$\mathbf{e}_1 = \mathbf{l}$$

is the (unit) *tangent*, $\mathbf{e}_2$ is the *principal normal*, and $\mathbf{e}_3$ is the *second normal* of this congruence. Vector $\boldsymbol{\kappa} = \kappa \mathbf{e}_2$ is the *curvature vector* of the congruence and scalars $\kappa$ and $\tau$ of Eq. (3.14) are the *curvature* and *torsion* of the congruence, respectively. A Frenet moving frame defines (at least locally) three two-dimensional distributions of planes: $\pi(\mathbf{e}_1, \mathbf{e}_2)$-*osculating planes*, $\pi(\mathbf{e}_2, \mathbf{e}_3)$-*normal planes* and $\pi(\mathbf{e}_3, \mathbf{e}_1)$-*rectifying planes* [6]. It follows from Eqs. (3.9), (3.12) and (3.15) that

(3.16) $$\mathbf{b} = b_{(l)} \mathbf{l} + b_{(m)} \mathbf{m},$$

and

(3.17) $\quad\mathbf{e}_1 = \mathbf{l}, \quad \mathbf{e}_2 = \cos\vartheta\,\mathbf{m} + \sin\vartheta\,\mathbf{n}, \quad \mathbf{e}_3 = -\sin\vartheta\,\mathbf{m} + \cos\vartheta\,\mathbf{n}.$

Consequently, according to Eq. (3.14), we obtain that:

(3.18)
$$\nabla_{\mathbf{l}}^g \mathbf{b} = \left[\partial_1 b_{(l)} - b_{(m)}\kappa\cos\vartheta\right]\mathbf{l} + \left[\partial_1 b_{(m)} + b_{(l)}\kappa\cos\vartheta\right]\mathbf{m}$$
$$+ \left[b_{(m)}(\tau - \partial_1\vartheta) + b_{(l)}\kappa\sin\vartheta\right]\mathbf{n}, \quad \kappa > 0.$$

Therefore, at each body point, the local Burgers vector $\mathbf{b}$ of the congruence, as well as its variation $\nabla_{\mathbf{l}}^g \mathbf{b}$ in the $\mathbf{l}$ direction, are located in the same local slip plane normal to the $\mathbf{n}$ direction iff [5]

(3.19) $\quad b_{(m)}(\tau - \partial_1\vartheta) + b_{(l)}\kappa\sin\vartheta = 0, \quad \kappa > 0.$

Note that, according to Eq. (3.10) with $\varphi_{b,1} = \pi/2$, the considered congruence consists of *edge* effective dislocation lines iff

(3.20) $\quad b_{(l)} = \mathbf{b}\cdot\mathbf{l} = 0, \quad b_{(m)} \neq 0.$

So, in this case, Eq. (3.19) reduces to the following representation of the torsion $\tau$:

(3.21) $\quad \tau = \partial_1\vartheta \geq 0.$

In the following, we will consider the congruence of effective mixed dislocation lines restricted by the above condition. This means that the *climb component* (see [6])

(3.22) $\quad \mathbf{n}\cdot\nabla_{\mathbf{l}}^g\mathbf{b} = b_{(l)}\kappa\sin\vartheta, \quad \mathbf{n}\cdot\mathbf{b} = 0,$

of the local Burgers vector variation is admitted. Next, let us observe that Eq. (3.17) can be rewritten in the following complex form:

(3.23) $\quad \mathbf{N} = \mathbf{m} + i\mathbf{n} = (\mathbf{e}_2 + i\mathbf{e}_3)e^{i\vartheta}, \quad \mathbf{l} = \mathbf{e}_1,$

where

(3.24) $\quad \mathbf{N}\cdot\mathbf{N} = \mathbf{l}\cdot\mathbf{N} = 0, \quad \mathbf{N}\cdot\mathbf{N}^* = 2, \quad \mathbf{l}\cdot\mathbf{l} = 1,$

and the asterisk denotes the complex conjugation. Introducing the complex variable $\psi$ of the form:

(3.25) $$\psi = \kappa e^{i\vartheta}, \quad \kappa > 0,$$

where $\kappa$ is the curvature of the congruence, and taking into account Eq. (3.21), we can rewrite the generalized formulas of Frenet (3.14) and Eq. (3.15) in terms of the Volterra moving frame $(\mathbf{l}, \mathbf{N})$ and the complex variable $\psi$:

(3.26) $$\kappa = \frac{1}{2}\left(\psi^* \mathbf{N} + \psi \mathbf{N}^*\right), \quad \nabla_{\mathbf{l}}^g \mathbf{N} = -\psi \mathbf{l}.$$

Let $\Phi = \Phi_t = \left(\mathbf{E}_a(\cdot, t)\right)$, $t \in I \subset \mathbb{R}^+$, be a time-dependent Bravais moving frame. The instantaneous metric tensor $\mathbf{g}_t = \mathbf{g}[\Phi_t]$ is defined then by Eq. (2.7). Namely, if $\Phi^* = \Phi^*_t = \left(\mathrm{E}^a(\cdot, t)\right)$ denotes the Bravais moving coframe dual to $\Phi_t$, and

(3.27) $$\mathbf{E}_a(p, t) = \underset{a}{e^A}(\mathrm{X}(p), t)\partial_{A|p}, \quad \mathrm{E}^a(p, t) = \overset{a}{e}_A(\mathrm{X}(p), t)d\mathrm{X}_p^A,$$

then $\mathbf{g} = \mathbf{g}(p, t) = \mathbf{g}_t(p)$, $(p, t) \in B \times I$, where

(3.28) $$\begin{aligned} \mathbf{g}_t(p) &= \mathbf{g}[\Phi_t](p) = \delta_{ab}\mathrm{E}^a(p, t) \otimes \mathrm{E}^b(p, t) \doteq \mathbf{g}_t(p(\mathrm{X})), \\ \mathbf{g}_t(p(\mathrm{X})) &= g_{AB}(\mathrm{X}, t) d\mathrm{X}^A \otimes d\mathrm{X}^B, \quad d\mathrm{X}^A \equiv d\mathrm{X}^A_{p(\mathrm{X})}, \\ g_{AB}(\mathrm{X}, t) &= \overset{a}{e}_A(\mathrm{X}, t)\overset{b}{e}_B(\mathrm{X}, t)\delta_{ab}, \quad \mathrm{X} = \mathrm{X}(p), \end{aligned}$$

and it is assumed that $p = p(\mathrm{X})$ iff $\mathrm{X} = \mathrm{X}(p)$. The instantaneous long-range distortion of the continuized dislocated Bravais crystal is characterized by the family $\mathbf{S}_t[\Phi] \equiv \mathbf{S}[\Phi_t]$, $t \in I$, of tensor fields dependent on time as a parameter:

(3.29) $$\begin{aligned} \mathbf{S}_t[\Phi](p(\mathrm{X})) &= \mathbf{S}[\Phi_t](p(\mathrm{X})) \\ &\doteq S_{ab}{}^c(X, t)\mathrm{E}^a(X, t) \otimes \mathrm{E}^b(X, t) \otimes \mathbf{E}_c(X, \mathrm{t}) \equiv \mathbf{S}_t(X). \end{aligned}$$

Thus, taking into account Eqs. (3.28) and (3.29), we can define the instantaneous dislocation density tensor by Eqs. (2.10), (2.11) and the instantaneous scalar density of dislocations by - Eq. (2.12).

The above defined Volterra and Frenet moving frames are time-dependent too. The time-dependent scalars $\kappa$ and $\tau$ of the generalized Frenet formulas (3.14) can be treated now as those that distinguish one class of congruences of moving effective dislocation lines from another. Consequently, the time-dependent complex version of Eqs. (3.21)-(3.28) of these formulas needs additional *kinematic equations* defining the evolution of curvature and torsion. A method of deriving such equations, based on the Frenet formulas for a single curve in the Euclidean space $\mathbb{R}^3$, has been formulated in order to describe the motion of a very thin, isolated vortex filament [15] (see also [16]) and it has been generalized in order to describe a congruence time-dependent curves in a Riemannian space [5]. Namely, putting

$$(3.30) \qquad \begin{aligned} \partial_t \mathbf{N} &= \omega_1 \mathbf{N} + \omega_2 \mathbf{N}^* + \omega \mathbf{l}, \\ \partial_t \mathbf{l} &= \omega_3 \mathbf{N} + \omega_4 \mathbf{N}^* + \omega_5 \mathbf{l}, \end{aligned}$$

and noting the relations of Eq. (3.24) and their partial derivatives with respect to time, we obtain

$$(3.31) \qquad \omega_1 = i\varsigma, \quad \omega_1 = \omega_2 = 0, \quad \omega_3 = -\omega^*/2, \quad \omega_4 = -\omega/2,$$

where $\omega$ and $\varsigma$ denote the complex and real scalar defined on $B \times I$, respectively. So, we have

$$(3.32) \qquad \begin{aligned} \partial_t \mathbf{l} &= -\frac{1}{2}\left(\omega^* \mathbf{N} + \omega \mathbf{N}^*\right), \\ \partial_t \mathbf{N} &= \omega \mathbf{l} + i\varsigma \mathbf{N}, \quad [\omega] = [\varsigma] = \mathrm{s}^{-1}. \end{aligned}$$

The system of equations is not closed. Therefore, some additional conditions are needed. For example, it follows from Eqs. (3.26) and (3.32), that the condition

$$(3.33) \qquad \partial_t \nabla_\mathbf{l}^g \mathbf{N} = \nabla_\mathbf{l}^g \left(\partial_t \mathbf{N}\right),$$

leads to the following *kinematic consistency equations*:

(3.34)
$$\partial_t \psi + \partial_1 \omega - i\varsigma\psi = 0,$$
$$\partial_t \varsigma = \frac{i}{2}\left(\omega\psi^* - \omega^*\psi\right) = \operatorname{Im}\left(\omega^*\psi\right),$$

and means that the equation

(3.35)
$$\partial_t \kappa = \nabla_1^g \left(\partial_t \mathbf{l}\right)$$

should be fulfilled. If the scalar $\omega$ is real, then Eq. (3.34) reduces to the following system of three real equations for four real variables $\kappa$, $\vartheta$, $\varsigma$ and $\omega$:

(3.36)
$$\partial_1 \varsigma = \omega\kappa \sin\vartheta, \qquad \partial_t \kappa + \cos\vartheta\, \partial_1 \omega = 0,$$
$$\kappa\left(\varsigma - \partial_t\vartheta\right) + \sin\vartheta\, \partial_1 \omega = 0,$$

where the versor **l** is treated as a fixed variable. It admits a broad class of nonlinear models of kinematics of the effective dislocation lines [5]. Particularly, if $\vartheta = \pi/2$, then Eq. (3.36) reduces to the relations

(3.37)
$$\partial_t \kappa = 0, \qquad \partial_1 \omega = \varsigma\kappa, \qquad \partial_1 \varsigma = -\omega\kappa,$$

admitting a *static congruence* of effective Volterra-type edge dislocation lines of torsion zero, being intersections of two orthogonal families of surfaces in $B_g$: crystal surfaces (on which the dislocations are located) and virtual slip surfaces [6].

## 4. Material flow

The Bravais moving frame $\Phi = (\mathbf{E}_a)$ defines a *plastic distortion tensor* **P** such that [1, 7]

(4.1)
$$\mathbf{E}_a = \mathbf{PC}_a, \qquad \mathbf{C}_a \mathbf{cC}_b = \delta_{ab},$$

where $C = (\mathbf{C}_a;\ a = 1, 2, 3)$ is a Cartesian basis defined on the Euclidean point space $E^3$ endowed with the Euclidean metric tensor $\mathbf{c}$. The Bravais moving coframe $\Phi^* = (\mathbf{E}^a)$ can be represented by a vectorial basis $\Phi_g^* = (\mathbf{E}^a;\ a = 1, 2, 3)$ of the linear module $W(B)$ (see [1], Appendix) uniquely defined by the following conditions:

(4.2) $$\langle \mathbf{E}^a, \mathbf{E}_b \rangle = \mathbf{E}^a \cdot \mathbf{E}_b \equiv \mathbf{E}^a \mathbf{g} \mathbf{E}_b = \delta^a_b, \qquad a, b = 1, 2, 3.$$

Then

(4.3) $$\mathbf{E}^a = \mathbf{P}^* \mathbf{C}^a, \qquad \mathbf{P}^* = (\mathbf{P}^{-1})^T, \qquad \mathbf{C}^a \mathbf{c} \mathbf{C}_b = \delta^a_b,$$

and the intrinsic material metric tensor can be written in the form [1]

(4.4) $$\mathbf{g} = \mathbf{P}^* \mathbf{c} \mathbf{P}^{-1} = \delta_{ab} \mathbf{E}^a \otimes \mathbf{E}^b.$$

Now, let us consider a time-dependent Bravais moving frame (Section 3). The partial derivative $\partial_t$ with respect to the time parameter is also designated further on, for simplicity, by a dot over letters. For example, it follows from Eqs. (3.27) and (4.1)-(4.3) that

(4.5) $$\begin{aligned} \dot{\mathbf{E}}_a(X, t) &= \partial_t e^A_a(X, t) \partial_A = \mathbf{S}_p(X, t) \mathbf{E}_a(X, t), \\ \mathbf{S}_p(X, t) &= \dot{\mathbf{P}}(X, t) \mathbf{P}(X, t)^{-1} = S^A{}_B(X, t) \partial_A \otimes dX^B, \\ \dot{\mathbf{P}}(X, t) &= \partial_t P^A{}_B(X, t) \partial_A \otimes dX^B. \end{aligned}$$

Thus, taking into account Eqs. (3.28) and (4.3)-(4.5), we obtain the following relations:

(4.6) $$\begin{aligned} \dot{\mathbf{g}} &= -2\mathbf{D}_p, \qquad \mathbf{D}_p = \frac{1}{2}\left( \mathbf{L}_p + \mathbf{L}_p^T \right), \\ \mathbf{L}_p &= S_{AB} dX^A \otimes dX^B, \qquad S_{AB} = g_{AC} S^C{}_B. \end{aligned}$$

Elastic behavior of matter is usually classified as reversible, inelastic behavior as irreversible. Like in thermodynamics, irreversibility in mechanics is much more involved than reversibility. Therefore it has not been possible to develop a unified the-ory which covers all the diverse inelastic phenomena. Nevertheless, it is a distinctive feature of inelasticity that the decisive motion processes occur in the interior of the bodies [17]. Thus, let us consider a smooth mapping $\chi : B \times I \to B$, $I = \langle 0, T \rangle$. The mapping is called a *material flow* if for each $t \in I$, the mapping $\chi_t(\cdot) = \chi(\cdot, t)$ is a local diffeomorphism $\chi_t : B_0 \to B_t = \chi_t(B_0) \subset B$ such that $\chi_0 = \mathrm{id}_B$. If $\xi = (\xi^a; a = 1, 2, 3)$ and $X = (X^A; A = 1, 2, 3)$ are two coordinate systems on $B$ and $\xi$ is treated as a reference coordinate system defined on $B_0$, then we can consider $X$ as a *convective Lagrange coordinate system* $X^A = \chi^A(\xi, t)$, $A = 1, 2, 3$, defined on $B$ at each instant $t \in I$ and such that for each $q \in B_0$, the following relations hold [7]:

$$(4.7) \qquad \chi^A(\xi(q), t) = X^A(\chi_t(q)), \qquad X^A(\chi_0(q)) = \delta^A_a \xi^a(q),$$

where

$$(4.8) \qquad \chi^A(\xi, t) = \chi^A_t(\xi), \qquad \chi^A_t = X^A \circ \chi_t \circ \xi^{-1} : \mathbb{R}^3 \to \mathbb{R},$$

denotes the (local) coordinate description of mappings $\chi_t$, $t \in I$. The mapping $\chi$ defines an intrinsic *material velocity* field $\mathbf{v}$ on $B$ by

$$(4.9) \qquad \begin{aligned} \mathbf{v}(p, t) &= \mathbf{v}_t(p) \in T_p B, & p \in \chi_t(U) \subset B_t, \\ \mathbf{v}_t &= \mathbf{V}_t \circ \chi_t^{-1}, & t \in I, \\ \mathbf{V}_t(q) &= \dot{\varphi}_q(t) \in T_{\chi_t(q)} B, & \varphi_q(t) = \chi(q, t), \quad q \in U \subset B_0, \end{aligned}$$

where $\dot{\varphi}_q$ denotes the vector field tangent to the curve $\varphi_q : I \to B$ and $(U, \xi)$ is a coordinate system on $B_0$. In the coordinate description of Eqs. (4.7) and (4.8) of $\chi$ we have [7]:

$$(4.10) \qquad \begin{aligned} \mathbf{v}(p, t) &= v^A(X^B(p), t) \partial_{A|p}, & p \in \chi_t(U), \\ v^A(\chi(\xi, t), t) &= \partial_t \chi^A(\xi, t), & [v^A] = \mathrm{cms}^{-1}. \end{aligned}$$

Let us consider the intrinsic counterpart **G** of the so-called *right Cauchy-Green tensor* defined as the time-dependent metric **g** pulled back by $\chi_t$, $t \in I$, that is [7, 18]:

$$(4.11) \quad \begin{aligned} \mathbf{G}(q,t) &= \chi_t^* \mathbf{g}_t(q) = G_{ab}(\xi(q),t) d\xi_q^a \otimes d\xi_q^b, \\ G_{ab} &= \chi^A{}_a \chi^B{}_b g_{AB} \circ \chi, \qquad \chi^A{}_a = \partial_a \chi^A, \qquad q \in B_0. \end{aligned}$$

The corresponding material volume 3-form is given by [18]:

$$(4.12) \quad \begin{aligned} \omega_G &= \chi_t^* \omega_g = \sqrt{G} d\xi^1 \wedge d\xi^2 \wedge d\xi^2, \\ G &= \det(G_{ab}) = J^2 g, \qquad J = \det(\chi^A{}_a), \end{aligned}$$

where

$$(4.13) \quad \begin{aligned} \omega_g &= \omega_g(p,t) = \omega_t(p), \qquad p \in \chi_t(B_0), \qquad t \in I, \\ \omega_g &= E^1 \wedge E^2 \wedge E^3 = \frac{1}{6} e_{ABC} dX^A \wedge dX^B \wedge dX^C, \end{aligned}$$

$e_{ABC} = e \varepsilon_{ABC}$, $\varepsilon_{ABC}$ denotes the permutation symbol associated with the cobase fields $dX^A$, $A = 1,2,3$, and Eqs. (2.12) and (3.28) were taken into account. The *intrinsic plastic strain tensor* $\mathbf{E}_p$ is given by

$$(4.14) \quad \mathbf{E}_p = \frac{1}{2}(\mathbf{G} - \mathbf{g}_0).$$

Let $L'_\mathbf{v} = \partial_t + L_\mathbf{v}$ denote an extended Lie differentiation operator [18]. Then (see Appendix)

$$(4.15) \quad \begin{aligned} L'_\mathbf{v} \mathbf{g} &= \dot{\mathbf{g}} + L_\mathbf{v} \mathbf{g}, \qquad L_\mathbf{v} \mathbf{g} = 2\mathbf{D}_g, \qquad \dot{\mathbf{g}} = \partial_t \mathbf{g}, \\ \mathbf{D}_g &= D_{AB} dX^A \otimes dX^B, \qquad D_{AB} = \frac{1}{2}\left(\nabla^g_A v_B + \nabla^g_B v_A\right), \end{aligned}$$

where $\mathbf{D}_g$ is called the *intrinsic rate of stretchings tensor* [7], and

$$(4.16) \quad \begin{aligned} L'_\mathbf{v} \omega_g &= \partial_t \omega_g + L_\mathbf{v} \omega_g, \\ L_\mathbf{v} \omega_g &= (\mathrm{div}_g \mathbf{v}) \omega_g, \qquad \partial_t \omega_g = \left(\partial_t \ln \sqrt{g}\right) \omega_g. \end{aligned}$$

Moreover

(4.17) $$\text{tr}\mathbf{D}_g = \text{div}_g \mathbf{v},$$

and [18]

(4.18) $$\chi_t^*\left(L'\mathbf{g}\right) = \partial_t \mathbf{G}, \quad \chi_t^*\left(L'_v \omega_g\right) = \partial_t \omega_g.$$

Let **P** denote the time-dependent *plastic distortion tensor* defined by Eqs. (3.27), (4.1) and let (see Eq. (4.5)):

(4.19) $$\dot{\mathbf{E}}_a = \mathbf{S}_p \mathbf{E}_a.$$

It follows from Eqs. (4.6) and (4.15)-(4.19) that

(4.20) $$\frac{1}{2} L'_v \mathbf{g} = \mathbf{D}_g - \mathbf{D}_p,$$

and, according to Eq. (4.14), we have

(4.21) $$\dot{\mathbf{E}}_p = \frac{1}{2}\dot{\mathbf{G}} = \chi_t^*\left(\mathbf{D}_g - \mathbf{D}_p\right).$$

It follows from Eq. (4.21) that if

(4.22) $$\forall (q, t) \in B_0 \times I, \quad \dot{\mathbf{E}}_p(q,t) = \mathbf{0},$$

then, since $\chi_0 = \text{id}_B$, we have

(4.23) $$\forall (q, t) \in B_0 \times I, \quad \chi_t^* \mathbf{g}_t(q) = \mathbf{g}_0(q),$$

and it should be

(4.24) $$\mathbf{D}_p = \mathbf{D}_g.$$

Moreover, it follows from Eqs. (4.6) and (4.16)-(4.18) that then

(4.25) $$\dot{\mathbf{g}} = -2\mathbf{D}_g$$

and

(4.26) $$\partial_t \ln \sqrt{g} + \text{div}_g \mathbf{v} = 0.$$

If the conditions (4.23) and (4.24) are fulfilled, then we will say that the material flow $\chi: B \times I \to B$ is *consistent* with the time-dependent internal length measurement (represented by the instantaneous material spaces $B_t$, $t \in I$). Because

(4.27)
$$E^a = (\chi_t^{-1})_* E_0^a,$$
$$(\chi_t^{-1})_* : W^*(B_0) \to W^*(B_t),$$

where $\Phi_0 = (E_0^a)$, $E_0^a = E^a(\cdot, 0)$, is a Bravais moving coframe considered at the instant $t = 0$ and $(\chi_t^{-1})_*$ is the *tangent mapping* acting according to the following formulae:

(4.28)
$$E^a = \overset{a}{e}_A dX^A, \qquad E_0^a = \overset{a}{e}_{0,c} d\xi^c,$$
$$\overset{a}{e}_A = \chi_{t,A}^c \overset{a}{e}_{0,c} \circ \chi_t^{-1}, \qquad (\chi_{t,A}^a \circ \chi_t)\chi_{t,b}^A = \delta_b^a,$$

the instantaneous metric tensor $\mathbf{g}_t$ is an intrinsic counterpart of the so-called *left Cauchy-Green tensor* defined as the push-forward of $\mathbf{g}_0$ by $\chi_t$, $t \in I$, that is [18]:

(4.29)
$$\mathbf{g}_t = (\chi_t^{-1})_* \mathbf{g}_0 = g_{t,AB} dX^A \otimes dX^B, \qquad \mathbf{g}_0 = g_{0,ab} d\xi^a \otimes d\xi^b,$$
$$g_{t,AB} = \chi_{t,A}^a \chi_{t,B}^b g_{0,ab} \circ \chi_t^{-1}, \qquad g_{0,ab} = \overset{c}{e}_{0,a} \overset{d}{e}_{0,b} \delta_{cd},$$

where $\mathbf{g}_0 = \mathbf{g}[\Phi_0]$ and Eqs. (4.7), (4.8) and (4.11) were taken into account. Moreover, if a material flow is consistent with the instantaneous internal length measurements, then it follows from Eq. (4.26) that the condition

(4.30)
$$\mathrm{div}_g \mathbf{v} = 0$$

is equivalent to the preservation of the body material volume in a rate-sensitive plastic regime. Thus, it is a counterpart of the *incompressibility condition* in the theory of perfectly plastic materials. A material flow consistent with the instantaneous internal length measurements and fulfilling the above incompressibility condition will be called the *conservative material flow*. Note that the above preservation of volume applies to the plastic motion only. However, it has been accepted in most macroscopic theories of elastoplasticity, that hydrostatic pressure and tension have a negligible influence: elastoplastic flow does not alter the density of the body [17]. This situation changes at very high pressure only. If we restrict ourselves to sufficiently low pressures, then there is no volume change by the total elastoplastic deformation, hence also no volume change by the elastic part of the total deformation [17].

## 5. Conservative material flows

Let us consider a Bravais moving frame $\Phi = (\mathbf{E}_a)$ and a Volterra moving frame $\Upsilon = (\mathbf{l}, \mathbf{m}, \mathbf{n})$ defined by Eqs. (3.9)-(3.15), and let us assume that these moving frames are time-dependent. It defines a time-dependent congruence $\mathbf{C}[\mathbf{l}]$ of mixed Volterra-type effective dislocation lines located in the time-dependent material space $B_g$. The local slip planes of this congruence are identical with the local crystal planes normal to the $\mathbf{E}_3$-direction of the Bravais moving frame $\Phi$ and that is why we will call them *local glide planes*. We will say that a material flow is *consistent* with the distri-bution $\pi_\mathbf{n}(\mathbf{l}, \mathbf{m})$ of local glide planes associated with the considered Volterra moving frame, if the corresponding intrinsic rate of stretchings tesor $\mathbf{D}_g$ is constrained by the following counterparts of kinematic conditions considered in the theory of perfect plasicity [7]: the material flow is conservative and the local glide planes are *instantaneously inextensible planes*. The last condition can be formulated as follows ([7], cf. [19]): if $\mathbf{u}$ is a vector field on $B_g$ such that

(5.1) $\quad\quad\quad\quad u_{(n)} = \mathbf{u} \cdot \mathbf{n} = 0, \quad\quad \text{i.e.,} \quad \mathbf{u} = u_{(l)}\mathbf{l} + u_{(m)}\mathbf{m},$

then

(5.2) $$\mathbf{u}\mathbf{D}_g\mathbf{u} = 0, \quad \text{i.e.,} \quad D_{AB}u^A u^B = 0.$$

The intrinsic material velocity field $\mathbf{v}$ of a material flow consistent in this sense with the Volterra moving frame is called the *dislocation flow velocity* [7]. The modulus $v_g = \|\mathbf{v}\|_g$ of this velocity will be called the *dislocation flow speed*. The components $D_{AB}$ of the intrinsic rate of stretchings tensor $\mathbf{D}_g$ corresponding to a dislocation flow velocity has the following representation [7]:

(5.3)
$$D_{AB} = \dot{\gamma}(S_A n_B + n_A S_B) = D(s_A n_B + n_A s_B), \quad s_A n^A = 0,$$
$$S_g = \sqrt{S_A S^A} = \sqrt{1+\delta_g^2}, \quad S_A = g_{AB}S^B = \delta_g l_A + m_A = S_g s_A,$$
$$\delta_g = D_{nl}/D_{nm}, \quad D_{nl} = \mathbf{n}\mathbf{D}_g\mathbf{l}, \quad \dot{\gamma} = D_{nm} = \mathbf{n}\mathbf{D}_g\mathbf{m}, \quad D = S_g \dot{\gamma},$$

where the versor $\mathbf{s} = s^A \partial_A$ is a Riemannian counterpart of the so-called *direction of shear* and $\dot{\gamma}$, $[\dot{\gamma}] = s^{-1}$, denotes the rate of inelastic shear in this direction. The pair $(\mathbf{s}, \mathbf{n})$ defines the *local slip system* with a resulting slip conditioned by a local gliding of Volterra-type effective dislocation lines in the $\mathbf{m}$-direction and by additional local slips along these lines (cf. [6]). Note that

(5.4) $$\cos \psi = \mathbf{s} \cdot \mathbf{m} = \frac{1}{S_g} > 0, \quad -\frac{\pi}{2} < \psi < \frac{\pi}{2}.$$

Let us consider the case of an *infinitesimally conformal equidistant material space* $B_g$ defined by the following condition [1, 20] (cf. [21]):

(5.5)
$$\nabla_A^g \varphi_B = \alpha g_{AB}, \quad 0 \neq \alpha \in C^\infty(B),$$
$$\varphi_A = \varphi_g n_A, \quad n_A n^A = 1, \quad \varphi_g > 0,$$

where $\nabla^g = (\Gamma_{BC}^A[\mathbf{g}])$ denotes the Levi-Civita covariant derivative with its Christoffel symbols $\Gamma_{BC}^A[\mathbf{g}]$ corresponding to the metric tensor $\mathbf{g}$ (dependent in general on time treated as a parameter). It follows from Eq. (5.5) that

$$(5.6) \quad \varphi_A = \partial_A \varphi, \quad \varphi \in C^\infty(B),$$
$$n^A \nabla^g_A n_B = 0,$$

and the space $B_g$ is foliated by the family $\Sigma = (\Sigma_c, c \in \mathbb{R})$ of surfaces of the form

$$(5.7) \quad \Sigma_c = \varphi^{-1}(c), \quad d\varphi \neq 0,$$

being, according to Eq. (3.13), crystal surfaces normal to the **n**-direction. Moreover, in this case for each $p \in B$ there exists a coordinate system $X = (X^A): U \to \mathbb{R}^3, p \in U, X^3 = \varphi|U$, such that

$$(5.8) \quad \Sigma_c = \{q \in U : X^3(q) = c\},$$

and Eq.(3.28) reduces to

$$(5.9) \quad \mathbf{g}(X, t) = \mathbf{g}_t(X) \doteq \Psi(X^3, t)\mathbf{a}(X^\kappa, t) + dX^3 \otimes dX^3,$$

where it was denoted

$$(5.10) \quad \Psi = a^2 e^{-2h}, \quad h \in C^\infty(\mathbb{R} \times I), \quad a \in C^\infty(I).$$

The time-dependent Bravais moving frame is, in the coordinate system of Eq. (5.8), given by

$$(5.11) \quad \mathbf{E}_\alpha(X^C(p), t) \doteq \Psi^{-1/2}(X^3(p), t)\mathbf{a}_\alpha(X^\kappa(p), t),$$
$$\mathbf{E}_3 \doteq \partial_3, \quad \alpha, \kappa = 1, 2,$$

where the vector fields $\mathbf{E}_a : U \times I \to W(B)$ have been identified with the vector fields $\mathbf{E}_a \circ X^{-1}: X(U) \times I \to W(B)$. The surfaces $\Sigma_{c,t} = (\Sigma_c, \mathbf{a}_{c,t}), t \in I$, where

$$(5.12) \quad \mathbf{a}_{c,t} = \Psi(c, t)\mathbf{a}_t, \quad \mathbf{a}_t = \mathbf{a}_t(X^\kappa) = \mathbf{a}(X^\kappa, t) = a_{\alpha\beta}(X^\kappa, t)dX^\alpha \otimes dX^\beta,$$

are then instantaneous *umbilical* crystal surfaces with the *constant mean curvatures* $H_c = H_c(t), (c, t) \in R \times I$, given by [1]

$$(5.13) \quad H_c(t) = H(c, t), \quad H = \partial_3 h.$$

Further on we will assume that

(5.14) $$\forall t \in I, \ \Psi(0, t) = 1.$$

The Christoffel symbols defined by the metric tensor of Eq. (5.9) are given by [20]

(5.15) $$\begin{aligned} \Gamma^3_{33}[\mathbf{g}] &= \Gamma^3_{3\alpha}[\mathbf{g}] = \Gamma^\alpha_{33}[\mathbf{g}] = 0, \\ \Gamma^\alpha_{\beta 3}[\mathbf{g}] &= -H\delta^\alpha_\beta, \qquad \Gamma^3_{\alpha\beta}[\mathbf{g}] = Hg_{\alpha\beta}, \\ \Gamma^\kappa_{\alpha\beta}[\mathbf{g}] &= \Gamma^\kappa_{\alpha\beta}[\mathbf{a}_c] = \Gamma^\kappa_{\alpha\beta}[\mathbf{a}], \end{aligned}$$

where $\nabla^a = \left(\Gamma^\kappa_{\alpha\beta}[\mathbf{a}]\right)$ is the Levi-Civita covariant derivative based on the metric tensor $\mathbf{a}$ and the dependence on the temporal parameter has been omitted.

Since

(5.16) $$L_{AB} = \nabla^g_B v_A = \partial_B v_A - \Gamma^C_{BA}[\mathbf{g}] v_C,$$

it follows from Eqs. (4.15) and (5.9)-(5.15) that if

(5.17) $$\mathbf{v}(X, t) = v^\alpha(X, t)\partial_\alpha, \qquad v^3 = v_{(n)} = \mathbf{v} \cdot \mathbf{n} = 0,$$

then

(5.18) $$\begin{aligned} L_{\alpha 3} &= Hv_\alpha, \qquad L_{3\alpha} = \partial_3 v_\alpha + Hv_\alpha, \qquad L_{33} = 0, \\ L_{\alpha\beta} &= \partial_\beta v_\alpha - \Gamma^\kappa_{\beta\alpha}[\mathbf{a}] v_\kappa = \Psi \nabla^a_\beta \bar{v}_\alpha, \qquad v_\alpha = \Psi \bar{v}_\alpha, \end{aligned}$$

where

(5.19) $$v_\alpha = g_{\alpha\beta} v^\beta = \Psi \bar{v}_\alpha, \qquad \bar{v}_\alpha = a_{\alpha\beta} v^\beta,$$

and, in general, $\bar{v}_\alpha = \bar{v}_\alpha(X^\kappa, X^3, t)$, $\alpha, \beta, \kappa = 1, 2$. Thus

(5.20) $$\begin{aligned} D_{\alpha 3} &= \tfrac{1}{2}\partial_3 v_\alpha + Hv_\alpha, \qquad D_{33} = 0, \\ D_{\alpha\beta} &= \Psi \bar{D}_{\alpha\beta}, \qquad \bar{D}_{\alpha\beta} = \tfrac{1}{2}\left(\nabla^a_\alpha \bar{v}_\beta + \nabla^a_\beta \bar{v}_\alpha\right). \end{aligned}$$

If $\mathbf{v}$ of Eq. (5.17) is the dislocation flow velocity, then it follows from Eqs. (5.3), (5.11), (5.19) and (5.20) that the following relations hold:

(5.21) $$D_{\alpha\beta} = 0, \quad \text{i.e.,} \quad \nabla^a_\alpha \bar{v}_\beta + \nabla^a_\beta \bar{v}_\alpha = 0,$$

and

(5.22) $$D_{\alpha 3} = Ds_\alpha, \quad \text{i.e.,} \quad \frac{1}{2}\partial_3 v_\alpha + Hv_\alpha = \dot{\gamma} S_g s_\alpha.$$

Eq. (4.30) follows now from Eqs. (4.17), (5.9) and (5.21).

It follows from Eqs. (5.8), (5.9) and (5.21) that the dislocation flow velocity **v** defines (up to its dimension) a Killing vector field (e.g. [14], [22]) for the crystal surfaces $X^3 = c$, $c \in \mathbb{R}$, being here virtually glide surfaces [1]. Note that the so defined Killing vector fields can act as the so-called *scalar preserving isometries* which leave invariant the intrinsic metric tensor **g** as well as the scalar $\varphi \in C^\infty(B)$ of Eq. (5.7) [1]. For example, if the considered crystal surfaces $\Sigma_{c,t} = (\Sigma_c, \mathbf{a}_{c,t})$ have additionally constant Gaussian curvatures $K_c(t)$, $(c,t) \in \mathbb{R} \times I$, then we are dealing with the following classes of virtual glide surfaces [1]. If $\Sigma_{c,t}$ is a *parabolic* surface ($K_c(t) = 0$), then it admits as its motion, in the small at least, the deformation of a Euclidean plane characterizing the *single glide* case (Section 1): planar rotations and translations [23]. In the *hyperbolic* case ($K_c(t) < 0$) we ought to take into account that the three-dimensional particular Lorentz group can be considered as a deformation of a Euclidean plane changing a square into a rhomb [24] (the so-called *pure shear*). The remaining three-dimensional Lorentz transformations are planar Euclidean rotations or their compositions with pure shearing. The case of *elliptic* glide surfaces ($K_c(t) > 0$) can be considered as the one corresponding to an elementary act of plasticity connected with the phenomenon of crystal fragmentation in the plastic yielding process and called rotational *plasticity* [25]. We see that Eq. (5.21) can be considered as the condition defining generators of a group of *conservative material flows* preserving glide surfaces.

## 6. Orowan-type kinematics

Let us return to the definition of congruences of Volterra-type effective dislocation lines given by Eqs. (3.2)-(3.13) and let us consider a congruence $C[\mathbf{l}]$ of principal (Volterra-type) effective dislocation lines defined by the condition [6]

(6.1) $$\boldsymbol{\gamma}\mathbf{l} = \gamma\mathbf{l}, \qquad \|\mathbf{l}\|_g = 1, \qquad \gamma \in \mathbb{R}.$$

It can be shown that if the conditions (5.9)-(5.14) are additionally fulfilled, then [6]

(6.2) $$\boldsymbol{\gamma} = \gamma\left(-\boldsymbol{\gamma}_1 \otimes \boldsymbol{\gamma}_1 + \boldsymbol{\gamma}_2 \otimes \boldsymbol{\gamma}_2\right), \qquad \gamma \geq 0,$$

and

(6.3) $$\begin{aligned}\mathbf{t} &= t^a \mathbf{E}_a = 2\left(-\gamma\boldsymbol{\gamma}_3 + H\mathbf{E}_3\right), \\ t_g &= \|\mathbf{t}\|_g = 2\sqrt{\gamma^2 + H^2} > 0.\end{aligned}$$

where

(6.4) $$\begin{aligned}\boldsymbol{\gamma}_1 &= \frac{1}{\sqrt{2}}(\mathbf{k} + \mathbf{E}_3), \qquad \boldsymbol{\gamma}_2 = \frac{1}{\sqrt{2}}(\mathbf{k} - \mathbf{E}_3), \\ \boldsymbol{\gamma}_3 &= \cos\varphi\,\mathbf{E}_1 + \sin\varphi\,\mathbf{E}_2, \qquad \mathbf{k} = \sin\varphi\,\mathbf{E}_1 - \cos\varphi\,\mathbf{E}_2.\end{aligned}$$

It follows from Eqs. (6.2)-(6.4) that if $C[\mathbf{l}]$ is an arbitrarily chosen congruence of Volterra-type effective dislocation lines, then its local Burgers vector $\mathbf{b}$ is given by [6]:

(6.5) $$\begin{aligned}\rho\mathbf{b} &= -\gamma\left(\cos\varphi_{\mathbf{l},\mathbf{E}_3}\mathbf{k} + \cos\varphi_{\mathbf{k},\mathbf{l}}\mathbf{E}_3\right) + \mu\mathbf{m}, \\ \mu &= \frac{1}{2}t_g \sin\varphi_{\mathbf{l},\mathbf{t}}, \qquad \mathbf{l}\cdot\mathbf{m} = \mathbf{t}\cdot\mathbf{m} = 0, \qquad \|\mathbf{l}\|_g = \|\mathbf{m}\|_g = 1.\end{aligned}$$

For example, if

(6.6) $$\mathbf{l} = \boldsymbol{\gamma}_3,$$

then

(6.7) $$\rho\mathbf{b} = \mu\mathbf{m}, \quad \mathbf{m} = \mathbf{k}, \quad \mu = \sqrt{H^2 + \gamma^2}.$$

Note that if

(6.8) $$\gamma = 0,$$

then

(6.9) $$\rho b_g = H.$$

and, since the considered crystal surfaces are umbilical (Section 5), the normal curvature $\kappa_n$ of these surfaces is the same for all their tangent directions and [22]:

(6.10) $$\kappa_n = H.$$

It means that Eqs. (6.9) and (6.10) define a Riemannian counterpart of Eq. (1.2). Moreover, the considered effective edge dislocation lines can be interpreted as those describing a continuized Bravais crystal, endowed with a distribution of very small prismatic edge dislocation loops normal to the time-dependent **m**-direction [6].

If the direction of the dislocation flow velocity **v** of Eqs. (5.17)-(5.22) coincides with the direction of shear **s** of Eqs. (5.3) and (5.4), that is

(6.11) $$\mathbf{v} = v_g \mathbf{s}, \quad v_g > 0,$$

then

(6.12) $$\dot{\gamma} = \cos\psi \left( H v_g + \frac{1}{2} s^\alpha \partial_3 v_\alpha \right).$$

Thus

(6.13) $$\dot{\gamma} = H v_g \cos\psi,$$

if and only if

(6.14) $$s_\alpha \partial_3 s^\alpha = v_g^{-1} \partial_3 v_g.$$

Particularly, if Eqs. (6.9) and (6.13) are valid, then we obtain the following generalization of the Orowan kinematic relation (1.1):

(6.15) $$\dot{\gamma} = \cos\psi\, \rho b_g v_g,$$

where multiplier $\cos\psi$ is a counterpart of the so-called *directional coefficient* considered in the physical theory of plasticity [26]. Note that if additionally the direction of shear coincides with the direction of the local Burgers vector, that is

(6.16) $$\mathbf{v} = v_g \mathbf{m},$$

then in Eq. (5.3) we have $S_g = 1$ (or equivalently $\delta_g = 0$) and thus

(6.17) $$\mathbf{D}_g = D(\mathbf{m}\otimes\mathbf{n} + \mathbf{n}\otimes\mathbf{m}), \qquad D = \dot{\gamma} > 0.$$

In this case Eq. (6.14) reduces to

(6.18) $$m_\alpha \partial_3 m^\alpha = v_g^{-1} \partial_3 v_g,$$

and the Orowan kinematic relation takes the form of Eq. (1.1):

(6.19) $$\dot{\gamma} = \rho b_g v_g.$$

### 7. Final remarks

Let us consider a material flow $\chi(\cdot, t) = \chi_t$, $t \in I$, fulfilling the conditions (6.15)-(6.18) and consistent with the distribution of $\pi = \pi_\mathbf{n}(\mathbf{l}, \mathbf{m})$, $\mathbf{n} = \mathbf{E}_3 \doteq \partial_3$, of local glide planes being virtual slip surfaces for a congruence $C[\mathbf{l}]$ of effective edge dislocation lines defined by the condition

(7.1) $$\mathbf{l}\cdot\mathbf{E}_3 = 0,$$

and by the following form of its local Burgers vector:

$$\rho \mathbf{b} = H\mathbf{m}, \qquad (7.2)$$

where $H$ is the mean curvature of umbilical crystal surfaces (see Section 6). If $\mathbf{T}$ is a symmetric stress tensor defined on actual configurations $\chi_t(U) \subset B_t$, $t \in I$, of domains $U \subset B_0$, and identified with an *internal stress tensor* dependent on the distribution of dislocations and secondary point defects, then the scalar

$$\mathrm{T} = \mathbf{mTn}, \qquad [\mathrm{T}] = \mathrm{kgcm}^{-2}, \qquad (7.3)$$

can be interpreted as the field of *resolved shear stresses* acting in oriented local slip planes of the distribution $\pi$ in the direction $\mathbf{m}$ of the local Burgers vector $\mathbf{b}$ [7] (cf. [4]). There are various dislocation dynamics descriptions, treating T as driving stress of moving dislocations. For example, it has been experimentally established that at low temperatures, when the climb (see [6], Section 1) is negligible, a relationship between the dislocation flow speed $v_g$, interpreted as the mean dislocation speed in the presence of many secondary point defects, and the stresses, can be taken in the following form (see e.g. [3, 4] and [26]):

$$v_g = v_0 \left(\frac{\mathrm{T}}{\mathrm{T}_0}\right)^n, \qquad (7.4)$$

where $v_0$ is a characteristic velocity of the order of the elastic shear wave speed, and T is an effective resolved shear stress (represented here by the stress defined by Eq. (7.3)). The characteristic parameters $\mathrm{T}_0$ and $n$ may be, in general, dependent on the temperature and permanent strains. Moreover, we will assume that the following generalized version of the condition of non-negativeness of dissipation is fulfilled:

$$\mathrm{tr}(\mathbf{TD}_g) = 2\mathrm{TD} = 2\mathrm{T}\dot{\gamma} \geq 0, \qquad (7.5)$$

where Eqs. (6.17) and (7.3) were taken into account. It must be emphasized that Eq. (7.4) implies no physical interpretation of the mechanism of dislocation motion. Par-ticularly, it is not assumed that a critical value of stresses is needed for the activation of the dislocation motion (and thereby, to create conditions for the appearance of plastic deformation [26]).

If Eqs. (7.4) and (7.5) are admitted, then the condition

$$(7.6) \qquad m_\alpha \partial_3 m^\alpha = \frac{n}{T} \partial_3 T, \qquad T \geq 0,$$

should be fulfilled. In this case we obtain, according to Eqs. (6.9), (6.19) and (7.4), that

$$(7.7) \qquad \dot{\gamma} = \dot{\gamma}_0 \left( \frac{T}{T_0} \right)^n, \qquad T > 0,$$

where $\dot{\gamma}_0 = \dot{\gamma}_0(X^3, t)$ is a time-dependent characteristic local strain rate of the form:

$$(7.8) \qquad \dot{\gamma}_0(X^3, t) = H(X^3, t) v_0.$$

Finally, we see that the material flow defined by Eqs. (6.16), (6.17), (7.4) and (7.6)-(7.8) can be considered as the one consistent with the Orowan kinematic relation as well as with treatment of the resolved shear stresses as driving stresses of moving dislocations.

**Appendix**

In differential geometry is considered the so-called *Lie derivative* $L_\mathbf{u}$ with respect to the vector field $\mathbf{u} \in W(B)$ (see e.g. [11, 14] and [18]). For example, if we denote

$$(A.1) \qquad \mathbf{T} = T^A_{BC} \partial_A \otimes dX^B \otimes dX^C,$$

then the Lie derivative operator will act according to the rule:

$$(A.2) \qquad (L_{\mathbf{u}}\mathbf{T})^A_{BC} = u^D \partial_D T^A_{BC} - T^D_{BC}\partial_D u^A + T^A_{DC}\partial_B u^D + T^A_{BD}\partial_C u^D.$$

Particularly (see [1], Appendix):

$$(A.3) \qquad \begin{aligned} L_{\mathbf{u}}\partial_A &= -\partial_A u^B \partial_B, & L_{\mathbf{u}} dX^A &= \partial_B u^A dX^B, \\ L_{\mathbf{u}}\mathbf{v} &= [\mathbf{u}, \mathbf{v}], & L_{\mathbf{u}} f &= \mathbf{u}(f) = u^A \partial_A f, \end{aligned}$$

and

$$(A.4) \qquad L_{\mathbf{u}}\mathbf{g} = 2\boldsymbol{\varepsilon}, \qquad L_{\mathbf{u}} g^{1/2} = g^{1/2} \mathrm{div}_g \mathbf{u} = g^{1/2} \mathrm{tr}\boldsymbol{\varepsilon},$$

where (see [1], Appendix)

$$(A.5) \qquad \begin{aligned} \boldsymbol{\varepsilon} &= \varepsilon_{AB} dX^A \otimes dX^B, & \mathbf{u} &= u^A \partial_A, \\ \varepsilon_{AB} &= \frac{1}{2}\left(\nabla^g_A u_B + \nabla^g_B u_A\right), & u_A &= g_{AB} u^B, \end{aligned}$$

is a Riemannian counterpart of the so-called *small strain* considered in the continuum mechanics (e.g. [27]), and it was denoted

$$(A.6) \qquad \mathrm{div}_g \mathbf{u} = \nabla^g_A u^A = g^{-1/2} \partial_A \left(g^{1/2} u^A\right), \qquad \mathrm{tr}\boldsymbol{\varepsilon} = g^{AB}\varepsilon_{AB}.$$